\documentclass[a4paper,aps,prl,twocolumn,groupedaddress,showkeys,showpacs,
  floats,floatats,floatfix]{revtex4-1}
\usepackage[utf8]{inputenc}
\usepackage[english]{babel}
\usepackage{graphicx}
\usepackage{dcolumn} 
\usepackage{bm}
\usepackage{natbib}
\usepackage{latexsym}
\usepackage{mathrsfs}
\usepackage{amssymb}
\usepackage{amsmath}
\usepackage{amscd}
\usepackage{color}
\usepackage{pifont}
\usepackage{pstricks,pst-node,pst-text,pst-3d}
\usepackage{verbatim}
\usepackage[T1]{fontenc}
\newcommand{\RC}{{$\cal RC$}}
\newrgbcolor{Red}{1.0 0.0 1.0}
\begin{document}
\title{Temperature Resistant Optimal Ratchet Transport}
\author{C.~Manchein$^{1}$, A.~Celestino$^2$  and M.W.~Beims$^{2,3}$}
\affiliation{$^1$Departamento de F\'\i sica, Universidade do Estado de
  Santa Catarina, 89219-710 Joinville, Brazil} 
\affiliation{$^2$Departamento de F\'\i sica, Universidade Federal do Paran\'a,
         81531-980 Curitiba,  Brazil}
\affiliation{$^3$Max-Planck-Institute for the Physics of Complex Systems, N\"othnitzer Str.~38, 01187, Dresden, Germany, EU}
\date{\today}
%
\begin{abstract}
Stable periodic structures containing {\it optimal} ratchet transport, recently 
found in the parameter space dissipation {\it versus} ratchet parameter
[PRL {\bf 106}, 234101 (2011)], are shown to be resistant to reasonable 
temperatures, reinforcing the expectation that they are essential to explain 
the optimal ratchet transport in nature. Critical temperatures for their 
destruction, valid from the overdamping to close to the conservative limits, are 
obtained numerically and shown to be connected to the current efficiency,
given here analytically. Results are demonstrated for 
a discrete ratchet model and generalized to the Langevin equation with an 
additional external oscillating force.
\end{abstract}
\pacs{05.45.Ac,05.45.Pq}
\keywords{Shrimps, ratchet currents, optimal transport, dissipation} 
\maketitle

Ratchets systems are remarkable profitable due their ability to 
displace particles without an external bias, which is known as 
ratchet transport. One of the challenges in the large variety of ratchet 
phenomena, and experiments, is to unveil how large currents of 
particles can be attained and controlled. If such displacement of 
particles can be controlled in distinct ratchet devices, it should 
be of great validity for technological 
applications. In this context we mention some experiments which 
successfully observed ratchet transport,  namely cells to control 
cancer metastasis  \cite{campbell09}, solids and  drops using the 
Leidenfrost effect \cite{lagubeau11,Stout}, 
micro and nanofluids \cite{boom11}, particles in silicon membrane 
pores \cite{matthias03}, cold atoms \cite{carlo05}, among others.
In addition, ratchet models appear as natural candidates to explain
directed transport in biology \cite{julicher97}, and have been 
realized in a wide range of physical systems \cite{reimann02}. 
In the real-world, ratchet systems always suffer environmental 
effects, like temperature, for example, which give rise to new
current dynamics, and so need to be taken into account by any
comprehensive approach. Therefore, to explain such transport in nature, 
and for technological applications, it is of fundamental priority to 
understand the effect of noise (thermal or not) on the efficiency of 
the ratchet current. 

Recent works \cite{alan11-1,alan11-2} have shown that, in the 
absence of temperature, optimal ratchet currents (\RC s) occur 
along organized structures, called isoperiodic stable structures (ISSs), 
in parameter spaces which involve the asymmetry, amplitude, 
phase of the ratchet and dissipation. The ISSs should play a 
fundamental role on the current generation in nature, since they 
bound the optimal \RC\, region, delimit current reversal areas, 
and explain the remarkable diversity of possible \RC s.
Results have partially been extended to quantum systems 
\cite{carlo12}. The ISSs are Lyapunov stable islands 
and are supposed to be generic in 
dynamical systems \cite{jasonPRL93,grebogi93,bonattoR07,stoop10}. 
Thus, any attempt to understand general effects of the temperature 
on the \RC, should be performed by analyzing the behavior of the
ISSs, and their surroundings, when the temperature is taken into
account. This is the main goal of the present Letter: to show the 
transformations of the ISSs under temperature, and that 
optimal \RC s inside the ISSs are resistant to reasonable 
temperatures. This is the opposite of what is done in 
non-inertial thermal rocket ratchets, where \RC s are thermally 
activated. Our results also confirm that generic 
ISSs, like Shrimps discovered in deterministic dynamical systems 
\cite{jasonPRL93,grebogi93}, for example, persist to some noise 
intensity in stochastic systems. Up to our knowledge, there is only 
one work \cite{reimann07} which observed the tendency that currents 
(not \RC s) survive to reasonable noise effects inside ISSs.

A model which presents all essential features regarding unbiased 
\RC s, is the dissipative map \cite{carlo05}

\begin{eqnarray}
\left\{
\begin{array}{ll}
  p_{n+1} = \gamma p_n + K[\sin(x_n)+a\sin(2x_n+\phi)] + \xi(n), \cr
  x_{n+1} = x_n + p_{n+1},\cr
\end{array}
\right.
\label{map}
\end{eqnarray}
where $p_n$ is the momentum variable and $x_n$ de position, 
$n=1,2,\ldots,N$ represents the discrete time, $K$ is the nonlinearity 
parameter, and $\xi(n)$ is the stochastic variable obeying 
{ $\left<\xi(n)\right>=0$ and} $\left< \xi(n)^2\right>=2(1-\gamma)k_B T$, 
with $k_B=1$ being the Boltzmann constant and $T$ the temperature. For 
$T=0$ this model was 
considered for $K=6.5, a=0.5, 0\le \gamma<1$ in \cite{casatiPRL07}, and
in the parameter spaces combining pairwise all parameters
$\gamma,K,a,\phi$ \cite{alan11-1,alan11-2}. The dissipation 
parameter $\gamma$ reaches the overdamping limit for $\gamma=0$ and the 
conservative limit for $\gamma=1$. The ratchet effect appears due to the 
spatial asymmetry, which occurs with  $a\ne0$ and $\phi\ne m\pi$ 
($m=1,2,\ldots$), in addition to the time reversal asymmetry for 
$\gamma\ne1$. The above model is quite general to understand \RC s. 
The parameter $K$ represents the intensity of an external asymmetric 
force in space. 
But since for $\gamma=1$ the above model can be derived from a 
kicked Hamiltonian, $K$ is as well the external kick which drives 
the system out of equilibrium, { which is a necessary 
condition to obtain ratchet transport.}

\begin{figure}[htb]
  \centering
  \includegraphics*[width=0.6\columnwidth]{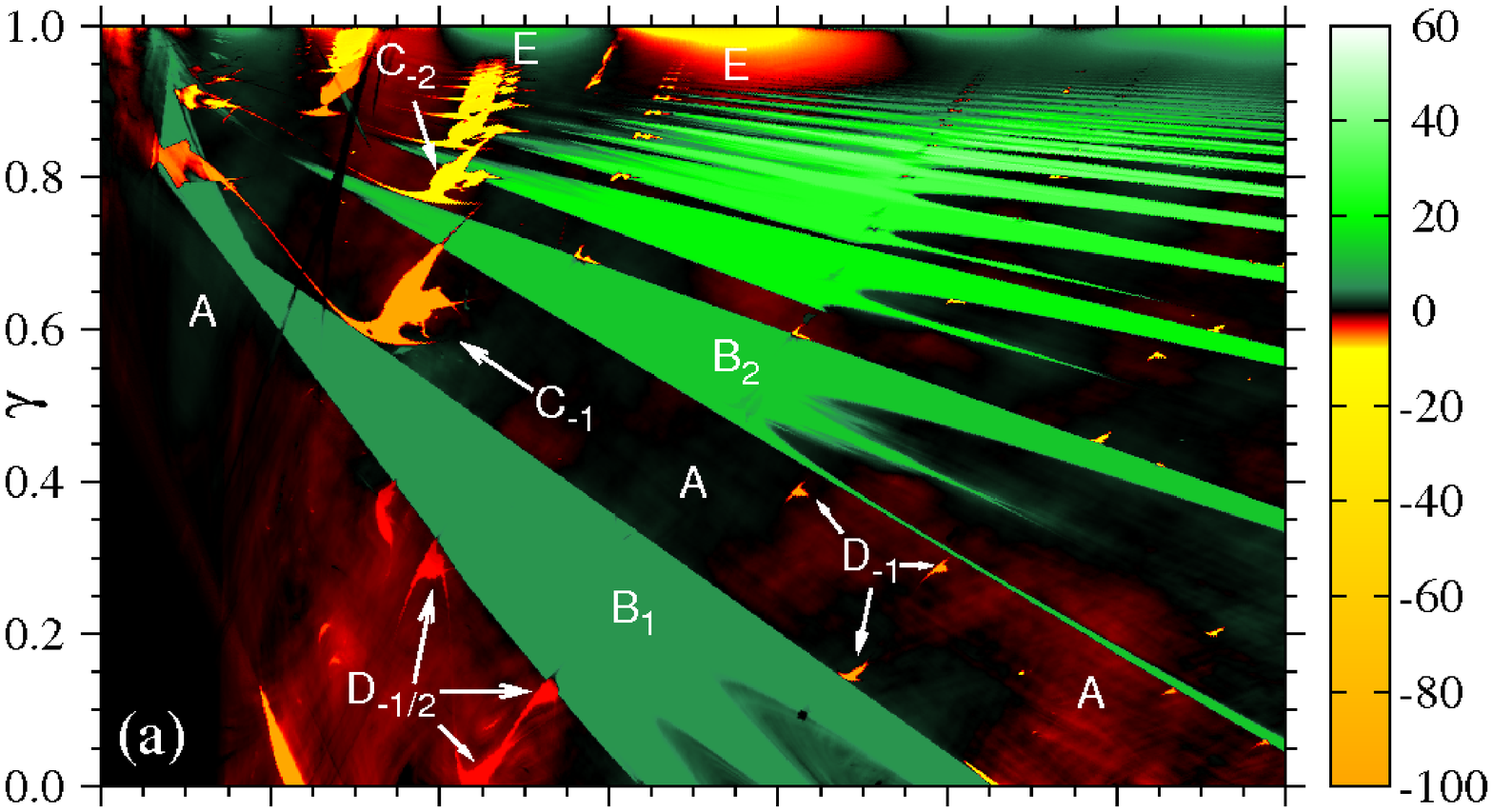}
  \includegraphics*[width=0.6\columnwidth]{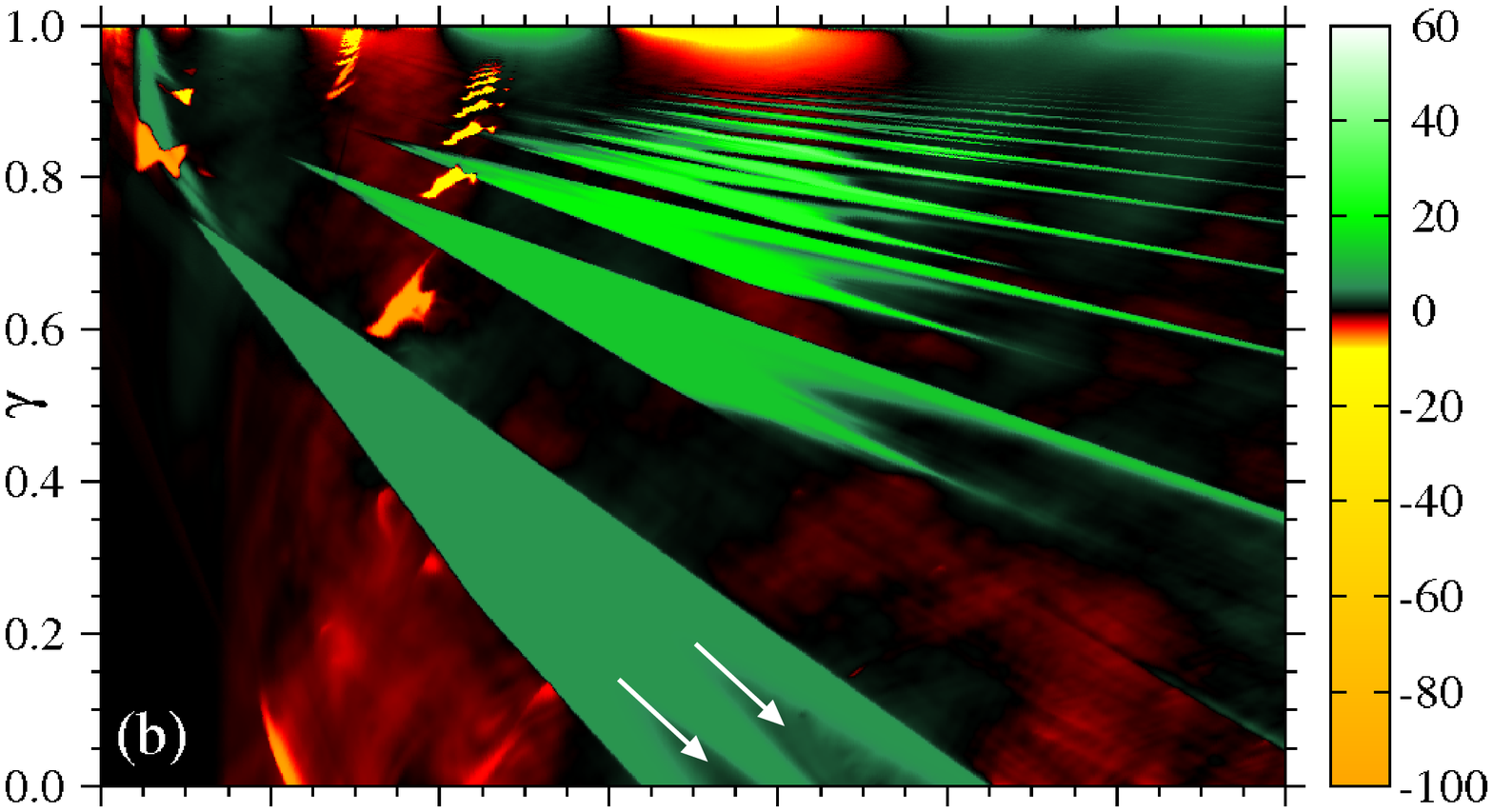}
  \includegraphics*[width=0.6\columnwidth]{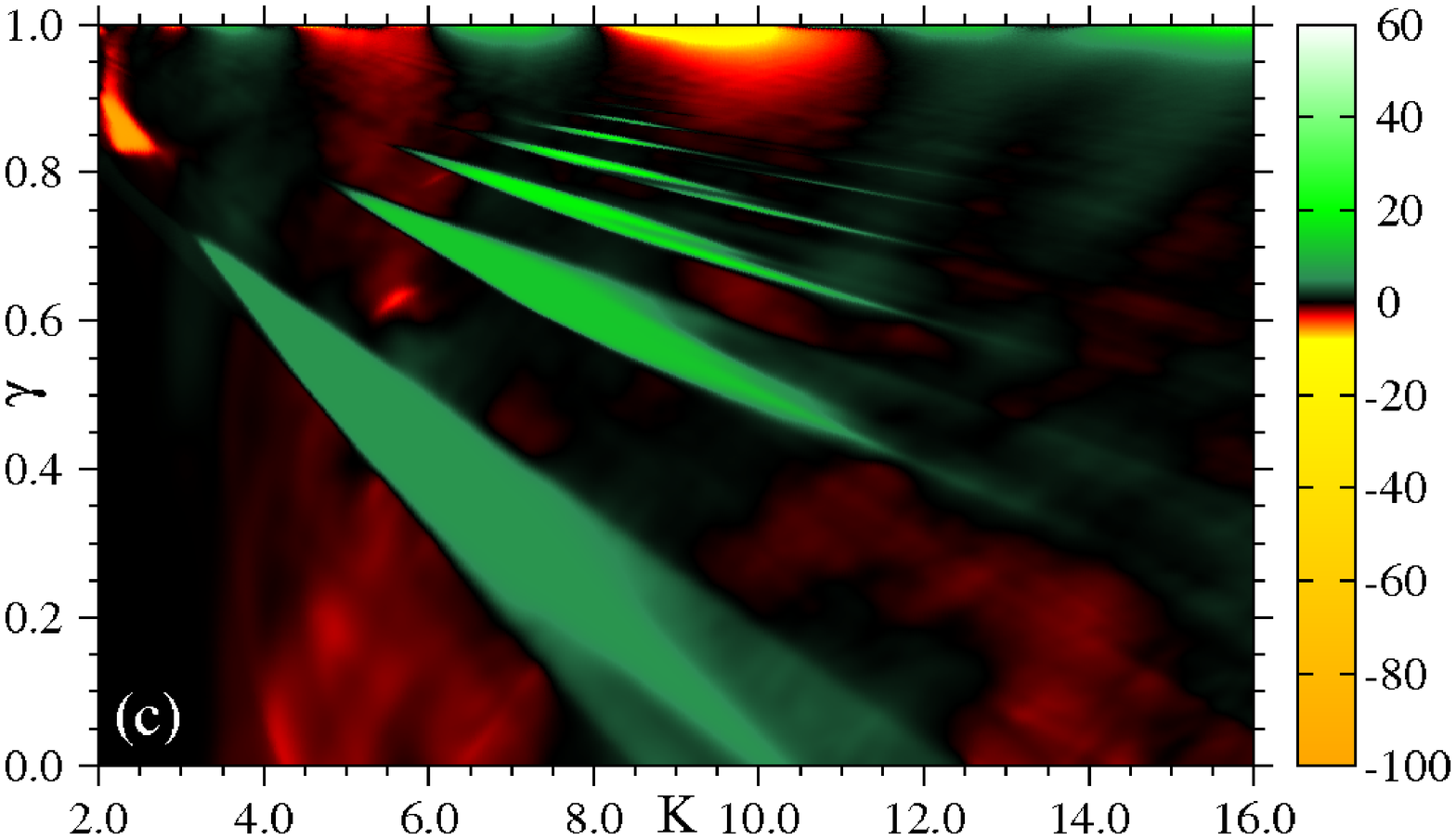}
 \caption{(Color online) The ${\cal RC}$ (see color bar) plotted in the 
parameter space ($K,\gamma$) with a   grid of $10^3\times 10^3$ points, 
$a=0.5$, $\phi=\pi/2$,  $M=10^4$ ICs with 
$\left<p_0\right>=\left<x_0\right>=0$ inside the unit cell ($-2\pi,2\pi$), 
$N=10^5,n_c=9\times 10^4$ iterations for (a) $T=10^{-5}$, 
(b) $T=10^{-3}$ and (c) $T=10^{-2}$}
  \label{PSS1}
\end{figure}
{For a small temperature $T=10^{-5}$, Fig.~\ref{PSS1}(a) shows the 
${\cal RC}=\frac{1}{M}\sum_{j=1}^M\left[\frac{1}{N-n_c}\sum_{n=n_c}^N p_n^{(j)}\right]$
(colors) as a function of $\gamma$ and $K$. $M$ is the number of initial 
conditions (ICs), $N$ is the total number of iterations used to calculate 
the \RC, and $n_c$ is obtained empirically and represents a 
cutoff value to avoid transient effects. The black color is related to 
small currents; dark green, green to white colors are related 
to increasing positive currents, while red to yellow colors related to 
increasing negative currents. Three main regions with distinct behaviors 
can be identified (for more details see Fig.~1 in \cite{alan11-1} for 
$T=0$): ($i$) a large  ``cloudy'' background $A$, mixed with black, dark 
green and red colors, showing a mixture of zero, small positive and negative  
currents. This is a chaotic region \cite{alan11-2} where the \RC\, is not that 
large, neither efficient, and is independent of the magnitude of the positive 
Lyapunov exponent; ($ii$) structures $B_L, C_L$ and $D_L$ with sharp borders 
(the ISSs) and distinct colors, which are embedded in the cloudy 
background region ($L$ is an integer or rational number). These ISSs 
are responsible for the larger \RC s and they align themselves along
preferred direction in the parameter space (see \cite{alan11-1,alan11-2}). 
Inside the ISSs the Lyapunov exponents (LEs) are zero and negative, and 
the motion is periodic, while a sequence of doubling 
bifurcations cascade occurs in a specific  direction. In the $C_L$ and 
$D_L$ ISSs, these bifurcations start from the inner part \cite{alan11-2}
of the ISSs, and reach the chaotic motion at specific borders. In the $B_L$ 
ISSs, saddle-node bifurcations of period-$1$ occur on the left and right 
borders, and were determined analitically in \cite{alan11-1,alan11-2}. 
Due to a crisis the analytical left borders do not match exactly with the 
numerical ones. A sequence of periodic doubling bifurcations 
occur when going down (smaller $\gamma$) to the right, where $K$ values 
increase [see white arrows in Fig.~\ref{PSS1}(b)]. It was also 
shown \cite{alan11-1,alan11-2} that the \RC\, is independent on the period 
of the orbits. The values $2\pi\,L$ determine the magnitude and direction of 
the \RC s. For example, the \RC\, inside the ISSs $B_L$, known as cuspidal 
singularities (cusps),  increases positively 
along the sequence $B_1,B_2,\ldots$. The \RC\, inside the $C_L$ ISSs, 
increases negatively along the sequence $C_{-1}, C_{-2},\ldots$. The 
smaller $D_L$ ISSs are the shrimp-shaped, well known to appear in 
many dynamical systems. In general, it was shown for 
$T=0$ \cite{alan11-2} that the \RC\,is always more efficient
inside the ISSs; ($iii$) the last region has strong positive and negative 
currents (region $E$), with not well defined borders, and occurs close to 
the conservative limit $\gamma=1$. This is related to a chaotic region 
(as shown in \cite{carlo05} for $T=0$), where the large \RC s are generated 
due to the accelerator modes which exist in the conservative limit. In 
Fig.~\ref{PSS1}(a) the temperature  $T=10^{-5}$ is small and all 
ISSs are mainly preserved. However, the ISSs have small elongated 
antennae, as seen in Fig.~1 from \cite{alan11-1},  which already 
disappeared for this small $T$. Consequently, the lines connecting the 
distinct $D_{-1}$, and  $D_{-1/2}$ ISSs, were destroyed by noise.  LEs 
inside these antennae become positive, meaning that the ISSs antennae 
are destroyed.

Figure \ref{PSS1}(b) shows the same parameter space, but now 
for a temperature  $T=10^{-3}$. Comparing with Fig.~\ref{PSS1}(a), 
we observe that the lower thin antennae from the $B_2,B_3,\ldots$
ISSs were destroyed. The $C_L$ ISSs begin to be destroyed from 
their borders, which are not sharply defined anymore. The $D_L$ 
shrimp-shaped ISSs have almost disappeared, and the optimal \RC s
inside them are substituted by the smaller \RC s from the chaotic
region.  The magnitude of the \RC s inside the remaining ISSs is 
essentially not affected by $T$, and still increases 
as the conservative limit is approached. All missing ISSs, or 
antennae, gave place to chaotic regions with smaller \RC s. In 
the chaotic regions $A$ and $E$, the \RC\, is almost unaffected 
by $T$.
 
As the temperature increases to $T=10^{-2}$, a significant amount
of ISSs start to be destroyed. This is shown in Fig.~\ref{PSS1}(c).
Almost all antennae from the $B_L$ ISSs are destroyed, remaining, 
in general, only the inner part of these structures, where the 
lowest period of the ISS is found. The less affected ISS is  $B_1$, 
which originally had the smallest \RC\, from the sequence 
$B_1, B_2, \ldots$. The boundaries of these ISSs are not sharp anymore,
but appear blurred. Remarkable is that the magnitude of the \RC s,
inside the inner part of the remaining $B_L$ ISSs, is almost the 
same. Along the period doubling bifurcations [see white arrows in  
Fig.~\ref{PSS1}(b)], $T$ destroys the optimal \RC, 
remaining just a region { (now chaotic)} with \RC s values a tiny 
amount larger when compared to the surrounding { background} chaotic 
region. Besides that, the $C_L$ ISSs are completely destroyed, giving 
place to the chaotic motion with almost zero current. Comparing to 
Fig.~\ref{PSS1}(a), in the chaotic regions $A$ and $E$ the \RC s 
remain unchanged.  Close to $K\sim 2.5$ and $\gamma\sim0.9$, we 
observe in Fig.~\ref{PSS1}(c) a region where 
larger \RC s occur. These larger \RC s were thermally activated, 
since they do not show up in Figs.~\ref{PSS1}(a)-(b). It gives 
an example, in distinction to  all other regions of the explored
parameter space, where the \RC\, is activated by $T$, instead
of being destroyed.

In this context the key question is: what is the critical $T$ to destroy 
the \RC\, in different stable regions (ISSs) of the parameter space? 
To answer this question the following simulation was performed. For each 
parameter combination ($K,\gamma$), $T$ was increased until the 
\RC\, was close to zero, within the precision $\Delta \sim\pm 3.0$, 
which  is a value bellow the minimum \RC\, allowed by the periodic 
constraints in ISSs with main periods $1$ and $2$ \cite{alan11-1},
that is, the main part of our parameter space. It is possible to observe 
that the close to zero \RC\, in the chaotic region, is independent of 
the parameters ($K,\gamma$) within the precision $\Delta$. This 
defines the critical temperature $T_c$, where the optimal \RC, due to 
periodic motion (ISSs), is transformed into the small \RC\, due to the 
chaotic motion. Figure \ref{Tc}(a)
\begin{figure}[htb]
  \centering
  \includegraphics*[width=0.7\columnwidth]{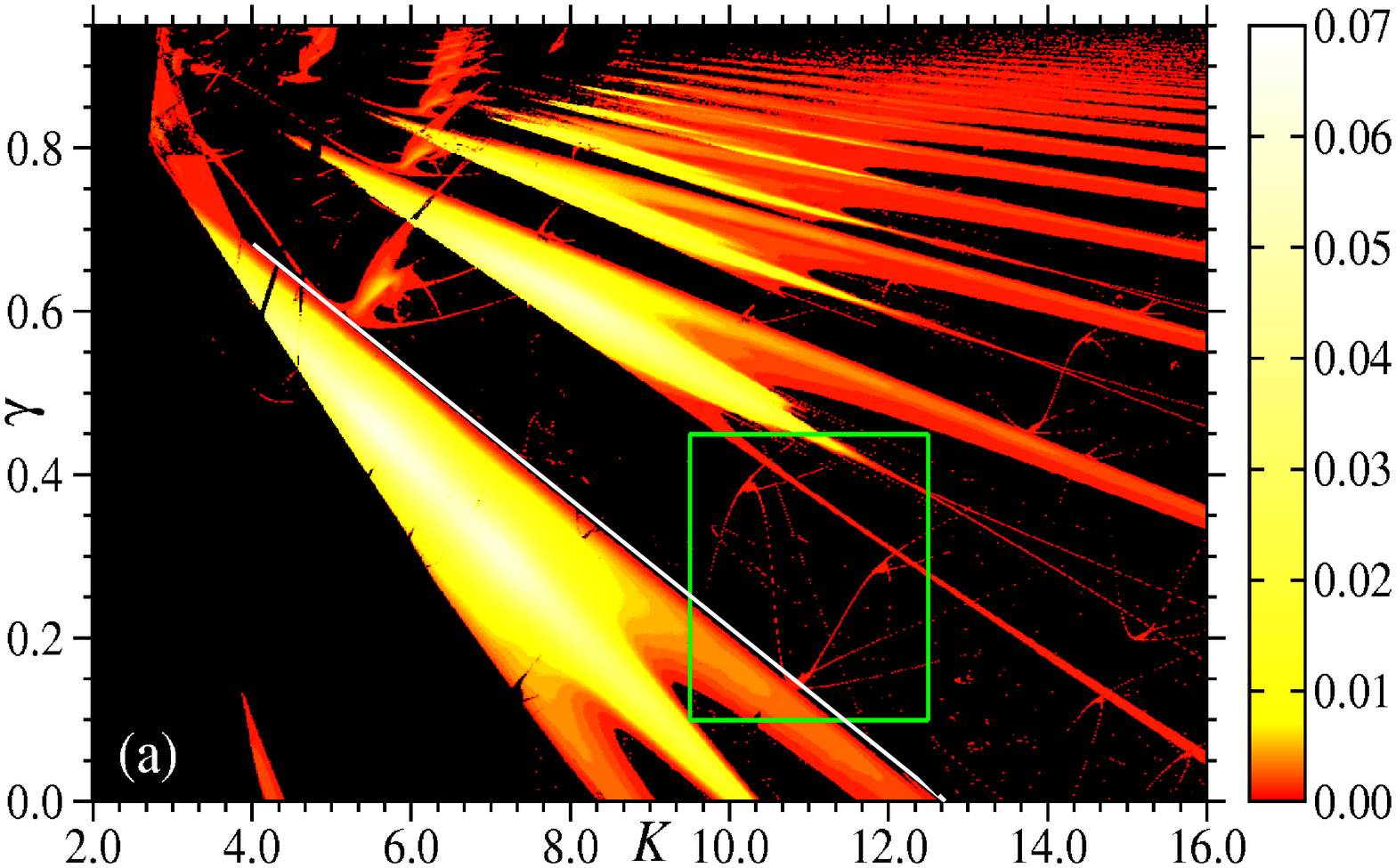}
  \includegraphics*[width=0.4\columnwidth]{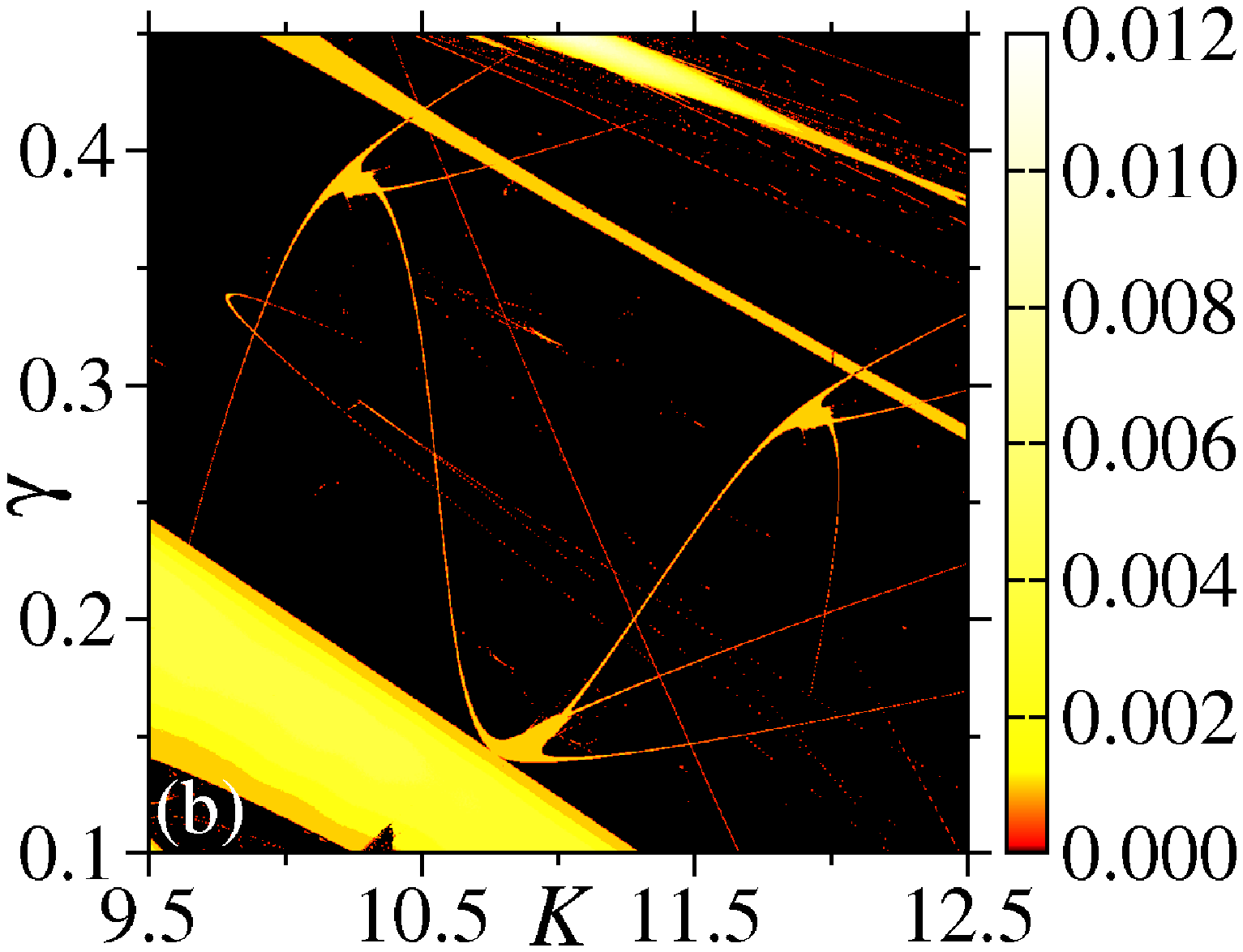}
  \includegraphics*[width=0.32\columnwidth]{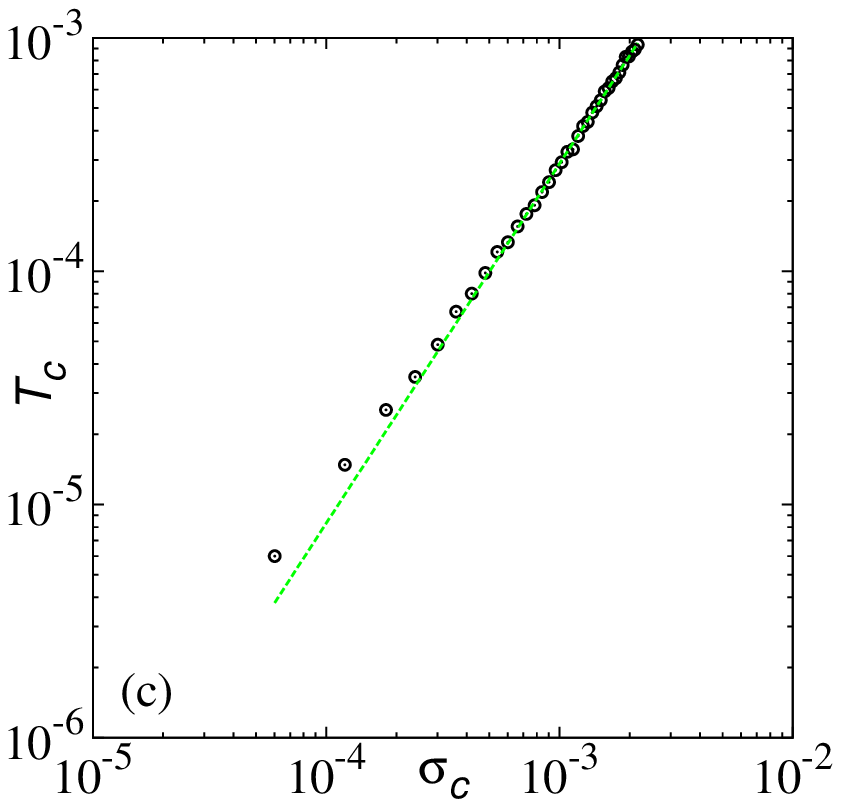}
 \caption{(Color online) The critical temperature $T_c$ (see color bar) 
necessary to destroy the optimal \RC\, plotted in the parameter space (a) 
from  Fig.~\ref{PSS1} and (b) inside the box from Fig.~\ref{Tc}(a) and
for  $a=0.5$, $\phi=\pi/2$ and same ICs from Fig.~\ref{PSS1}. 
{ In  (c) we show that $T_c\sim|\gamma-\gamma_c|^{1.5}$, for 
$K=2\pi$ and around $\gamma_c=0.5$.}}
  \label{Tc}
\end{figure}
shows $T_c$ (color bar) in the parameter space ($K,\gamma$). $T_c$ increases 
from black ($T_c\sim 0.0$) to red, yellow and 
white ($T_c\sim 0.07$). It is clearly observed that $T_c$ is larger inside 
the ISSs, especially inside their inner parts. Figure \ref{Tc}(b) shows in 
details $T_c$ from the box of Fig.~\ref{Tc}(a), with three connected 
shrimp-like structures. Thus, the ISSs can still be recognized in such 
$T_c$ analysis. Even though $T_c$ is larger inside the inner parts of 
the ISSs, it tends to decrease at their borders.

The physical origin of the thermally induced destruction of the optimal 
\RC\, can be rather complex. It depends very much on which region in the 
parameter space is considered. Inside the ISSs there is always at least 
one periodic stable attractor, which coexists with the unstable periodic 
orbits and chaotic repellers. With the introduction of $T$, the attracting
motion (periodic or not) becomes transient and the dynamics is ergodic, 
as earlier studies  about noise effects and transient chaos in dynamical 
systems indicate \cite{telbook}. Temperature also induces transitions between 
all possible dynamics: 
periodic/non-periodic attractors, chaotic repellers etc. Thus, for $T\ne0$, 
the final \RC s will be an average over the velocities of each of such dynamics, 
weighted by their lifetimes. Since the \RC\, is more efficient inside the 
ISSs \cite{alan11-2}, we should understand the effect of $T$ on the \RC s 
efficiency, which is defined by 
$\eta={\langle\langle p\rangle\rangle^2}/{\left|\langle\langle
p^2\rangle\rangle-D_0\right|}$ \cite{haenggi05}.
$D_0=k_BT$ is a measure of the thermal energy and 
$\langle\langle\ldots\rangle\rangle$ denotes an average over ICs, 
thermal realizations and time. For the ergodic case it reduces to 
just one average, over the asymptotic distribution.
Considering that the thermal ``kicks'' are typically weak enough,
so they can be neglected after one iteration, but rarely strong
enough to drive the orbit out of the attractor, then
it is possible to show that 

\begin{equation}
\eta=
\frac{\left(\sum_i\alpha_i\mbox{\RC}^{(i)}\right)^2}
{\left|\sum_i\alpha_i\langle p_{i,T=0}^2\rangle+(1-\gamma)D_0/(1+\gamma)\right|},
\label{eta}
\end{equation} 
where the sum is over all attractors, 
$\langle p_{i,T=0}^2\rangle$ is the mean of the momentum square on the attractor 
$i$ (for $T=0$), $\alpha_i=\mu_i\tau_i$ is the statistical weight of each attractor, 
where $\tau_i$ is the mean lifetime on attractor $i$, and 
$\mu_i=\lim_{n\to\infty}\langle m_i\rangle/n$, where $m_i$ is the number of 
times a given trajectory visited the attractor $i$ during the time $n$.
According to our simulations, transitions between deterministic dynamics
[the sums over attractors from Eq.~(\ref{eta})], induced by $T$, affect more 
the efficiency than pure thermal effects (proportional to $D_0$). Such 
transitions
may even increase the efficiency by allowing the orbit to access attractors
with higher currents related to the unstable dynamics from $T=0$, as could
be seen in Fig.~\ref{PSS1}(c). However, inside the ISSs the current is usually
optimal, thus $T$ tends to decrease its efficiency. Also, as $T_c$ is related
to very frequent transitions between different dynamics, the efficiency for
larger $T$ must decrease substantially.

As far as we have knowledge, there is no first principles theory that quantifies 
the critical temperatures which destroy the \RC s. Quasipotentials 
\cite{telbook} and escape rates \cite{reimann07} from periodic/non-periodic 
attractors and chaotic repellers, have been used to quantify $T_c$ in a context 
not related to \RC s. It is known \cite{telbook} that close to a saddle-node 
bifurcation, the dynamics of a higher-dimensional system can be reduced to an 
one-dimensional normal form,
and then the critical Gaussian noise follows $\sigma_c\sim|\gamma-\gamma_c|^{3/4}$,
with $\sigma_c$ being the critical noise amplitude. To check this, we plot $T_c$
in Fig.~\ref{Tc}(c) as a function of  $|\gamma-\gamma_c|$, for $K=2\pi$ and 
$\gamma_c=0.5$. This is a point at the border line from the $B_1$ ISS
(see white line in Fig.~\ref{Tc}(a)), which was calculated analytically in 
\cite{alan11-1,alan11-2}, and has a saddle-node bifurcation. Since in our case 
$\sigma\propto\sqrt{T}$ we expect that $T_c\sim|\gamma-\gamma_c|^{3/2}$. 
This is exactly what is observed in Fig.~\ref{Tc}(c) from our numerical simulation 
(cross points). For comparison, we plotted the exact continuous line 
$|\gamma-\gamma_c|^{3/2}$.

To generalize our results to a continuous, and more realistic system, 
we analyze the $T_c$ to destroy the optimal \RC\, in the Langevin 
equation: $\ddot x +\gamma\dot x - 5.0\left[\sin{(x)} 
+ 0.7\cos{(2x)}\right]-K_t\sin{(t)}+\xi(t)=0$, where 
$K_t$ is the amplitude of the external time oscillating force,
$\gamma$ is the viscosity, while the ratchet potential and the 
stochastic term $\xi(t) $ are like from Eq.~(\ref{map}). ISSs have 
been found \cite{alan11-1} for this problem for $T=0$ in the 
parameter space [$K_t,\chi=e^{(-\gamma)}$].
\begin{figure}[!ht]
   \includegraphics[width=0.7\columnwidth]{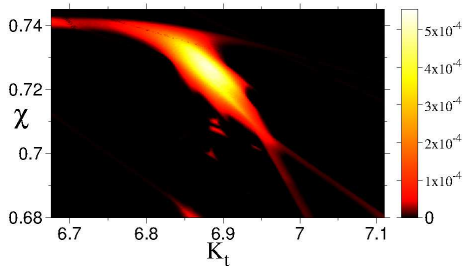}
  \caption{(Color online) The critical temperature $T_c$ in the 
parameter space ($K_t,\chi=e^{(-\gamma)}$) of the Langevin equation. 
The shrimp-like ISS can be identified.}
 \label{lang-Tc}
\end{figure}
Figure \ref{lang-Tc} shows the critical temperature $T_c$, required
to destroy the \RC\, in a portion of the parameter space  
($K_t,\chi$). The black color ($T_c=0$) means that no optimal \RC\, is 
observed, red, yellow to white colors are related to the increasing 
values of $T_c$ required to destroy the \RC. The shrimp-like 
ISS can be identified, showing that its inner part is more resistant 
(yellow) to external noise, while on its border (red), smaller noises 
are enough to destroy the optimal \RC s region.

Concluding,  optimal ISSs \RC s in the parameter space are shown to be 
resistant to reasonable temperatures $T$, i.e., it exists for thermal 
energies around $1000$ times smaller then the transport energy. Increasing 
the values of $T$, the \RC s inside the ISSs begin to be destroyed from 
their borders, and the 
ISSs start to become more and 
more blurred. We determined numerically the critical temperature $T_c$, 
required to destroy the optimal \RC s in the whole parameter space. 
Results show that, in general, as nearer of the ISSs' center, higher are
the critical temperatures. The ISSs persist in such a plot. Distinct 
regions in the parameter space present different dynamics and $T$ affects 
the optimal \RC s  distinguishable. The general physical explanation is 
that $T$ induces a transient 
chaos and also transitions between all possible dynamics: periodic and 
non-periodic attractors, chaotic repellers etc. An analytical expression
for the \RC\, efficiency was obtained, which combines all these dynamics.
It essentially shows that there are two thermal effects on the efficiency,
namely the slight variation in $\langle p^2\rangle$ inside the attractor, 
and the transitions between different dynamics. Considering the scale
of $T_c$, the first effect is practically neglectable and thus the \RC\,
is optimal inside the ISSs, where the transitions are very rare. For 
$T\ne 0$, the periodic attractor loses its dominance on the \RC, allowing 
the rise of those new kinds of dynamics that control the \RC\, above a 
critical value of $T$, $T_c$. This temperature is obtained numerically in 
Fig.~\ref{Tc} for the asymmetric standard map, in Fig.~\ref{lang-Tc}  for 
the Langevin equation with an external driving force, and is a direct 
measure of all complicated combined effects to destroy the \RC. 
It also gives quantitative information about the “global stability” of
attractors under noise in phase space. This allows us to relate the
robustness of the optimal RC to the global stability of solutions, rather
than to the local stability obtained from the Lyapunov analysis. As well
motivated by \cite{beale89}, information regarding global stability is 
usually more relevant in real-world systems where perturbations are finite. 
Hence,  our results underscore the fundamental role of isoperiodic stable 
structures in the real-world optimal ratchet transport, where environmental 
effects cannot be neglected. 

\vspace*{-0.5cm}


\end{document}